\documentclass[twocolumn,prl,aps,floatfix,showpacs]{revtex4}
\usepackage{graphicx}

\begin{document}
\title{Density functional theory description of hole-trapping in SiO$_2$: a successful self-interaction-corrected approach.}
\author{Mayeul d'Avezac}
\author{Matteo Calandra}
\author{Francesco Mauri}
\affiliation{Laboratoire de Min\'eralogie-Cristallographie, case 115, 4 place Jussieu, 75252, Paris cedex 05, France}
\date{\today}

\begin{abstract}
We present a  self-interaction-corrected (SIC)
density-functional-theory (DFT) approach for the description 
of systems with an unpaired electron or hole such
as spin 1/2 defect-centers in solids or radicals. 
Our functional  is easy-to-implement and its minimization
does not require additional computational effort with respect to 
ordinary DFT functionals. In particular it does not present
multi-minima, as the conventional SIC functionals.
We successfully validate the method studying
the hole self-trapping in quartz associated to 
the Al substitutional impurity. We show that our approach 
corrects for the well known failures of standard DFT functionals
in this system.
\end{abstract}
\pacs{71.15.Mb, 61.72.Bb, 71.55-i}

\maketitle
When an electron, a hole, or an electron-hole pair is introduced in an
insulator, its wavefunction can have different degrees of localization. It can
be localized (self-trapped) on a single atom/bond or it can be delocalized on a
larger scale.  Trapped electron or hole centers occur in very different
systems.  The F-centers in alkali halides are paramagnetic centers formed by
electron defects trapped on negative ion vacancies in the crystal.  They can be
obtained by introducing a stoichiometric excess of alkali metal atoms or
ionizing the crystal with radiation\cite{Mallia}.  Trapped hole defects are
found in alkaline earths oxides ([M]$^0$ centers)\cite{Lichanot}\cite{Schirmer}
upon substitution of the divalent alkaline earth atom with a monovalent one.

Several examples of self-trapping occur in silica.  Holes and electrons induced
by ionizing radiation self-trap in amorphous SiO$_2$\cite{Stoneham} but not in
quartz\cite{Haies}.  In quartz, localization of holes is achieved via the
substitution of Si with a trivalent atom (e.g. Al)\cite{Stokbro}.  Similarly,
electron  self-trapping is obtained substituting a Si with a pentavalent atom
(e.g. P).  Self-trapped holes or electrons in quartz can also be obtained by
substituting Si with Ge and by removing or adding an electron with ionizing
radiation.  Excitons self-trap both in amorphous Silica and in
quartz\cite{Stoneham}.

The study of self-trapping of centers in both amorphous and crystalline SiO$_2$ is
a subject of important technological implications.  The creation of
self-trapped centers in SiO$_2$ due to ionizing radiation or high intensity UV
light determines the degradation of UV-transmitting fibers(for e.g. UV
lithography). Moreover self-trapped defects are partly responsibles for the
failure rate of MOS devices which use amorphous SiO$_2$ as an insulating layer.

From a theoretical point of view, the description of trapped defects is
particularly challenging\cite{Stokbro} since a standard density functional
theory (DFT) approach based on the local spin density approximations (LSDA) or
on its improved version, the spin polarized generalized gradients
approximations (SPGGA), often fails to reproduce the localization of the defect
wavefunction.  Unrestricted Hartree Fock (UHF) calculations
\cite{Dovesi,Stokbro} usually correctly reproduces the self-trapping. However 
the UHF description of the defect-free system is usually less accurate than
that obtained with DFT. Also, UHF is computationally more demanding.

DFT predicts most F-centers in alkali halides to be delocalized\cite{Mallia}.
In the case of SiO$_2$, electron self-trapping is correctly described by DFT,
whereas holes are found delocalized\cite{Pacchioni, Stokbro}.  A prototype
example are neutral Al centers in SiO$_2$.  The structure of pure silica is
composed of corner-sharing [SiO$_{4}$] tetrahedrons. The Al doping occurs as
a substitution of a Si in the tetrahedral structure, resulting in consequential
modifications of electronic and geometric properties.  The Al-Si substitution
introduces in the system a hole which, according to electron spin resonance, is
localized on one of the surrounding oxygen.  On the contrary, and in clear
disagreement with experimental findings, DFT using standard functionals
predicts a hole wavefunction delocalized over the four surrounding
oxygens\cite{Pacchioni, Stokbro}. For this reason the description of hole
trapping on Al impurities in Silica has been defined  ``a challenge for
density functional theories''\cite{Stokbro}. 

As suggested by several authors \cite{Pacchioni,Stokbro} the failure of 
DFT in describing self-trapping in general, and self-trapping on Al centers in particular, 
might be due to the incomplete cancellation of the unpaired-electron
self-interaction.  This  cancellation occurs exactly in UHF. A possible
solution to the problem might be to use self-interaction corrected (SIC) 
functionals\cite{Perdew,Lundin}. It is indeed well known that SIC functionals
can describe electronic states which are not reproduced by LSDA/SPGGA
functionals\cite{Svane}.  Unfortunately, the implementation of DFT functionals
with self-interaction corrections on each orbital, besides being technically
complex, leads to multi-minima problems and sometimes degrades the results
obtained with standard functionals \cite{Goedecker}.

In this work we show that SIC functionals are indeed a remedy to the failure of
DFT in describing Al defects in Silica. We present an easy-to-implement SIC
functional for one-particle spin $1/2$ defects in solids.  In our SIC
functional approach we apply the self-interaction correction to the unpaired
electron only.  We treat the closed shell system formed by the remaining $2N$
electrons using standard DFT functionals but imposing the conditions that up
and down electrons with the same orbital quantum numbers have identical spatial
wavefunctions.  Our method is free from the multi-minima problem found in
previous SIC implementations and its computational cost is equal to that
of ordinary DFT functionals.  We validate the method against experimental data
by calculating optimized geometries and hyperfine coupling parameters of Al 
defects in silica.

We consider a system of $2N+1$ electrons with $N_{\uparrow}=N+1$ up electrons
and $N_{\downarrow}=N$ down electrons. The wavefunction of the $i^{th}$
$\sigma-$electron ($\sigma=\uparrow,\downarrow$) is $|\psi_{i\sigma}\rangle$.
Throughout the paper we assume the wavefunctions $|\psi_{i\sigma}\rangle $ to
be orthonormal.  We write $n_{\sigma}({\bf r})=\sum_{i=i}^{N_{\sigma}}
|\langle{\bf r}|\psi_{i\sigma}\rangle|^2$ and $n({\bf r})=\sum_{\sigma}
n_{\sigma}({\bf r})$.  The magnetization density is $m({\bf
r})=n_{\uparrow}({\bf r})- n_{\downarrow}({\bf r})$.

In spin polarized density functional theory the total energy functional is: 
\begin{eqnarray}
F[\{\psi_{i\sigma}\}]= T[\{\psi_{i\sigma}\}]
+E_{\rm KS}[n_{\uparrow},n_{\downarrow}]+ \nonumber \\
 + \sum_{\sigma={\uparrow,\downarrow}}\sum_{i,j=1}^{N_{\sigma}} \lambda_{ij}^{\sigma}
(\langle\psi_{i\sigma}|\psi_{j\sigma}
\rangle-\delta_{ij})
\end{eqnarray}
where $T[\{\psi_{i\sigma}\}] =
-\frac{1}{2}\sum_{\sigma}\sum_{i}^{N_\sigma}\langle\psi_{i\sigma}|\Delta|\psi_{i\sigma}\rangle$
is the single particle kinetic energy and $E_{\rm
KS}[n_{\uparrow},n_{\downarrow}]=E_{\rm ext}[n]+E_{\rm H}[n]+E_{\rm
xc}[n_{\uparrow},n_{\downarrow}]$. The functionals $E_{\rm ext}[n]$, $E_{\rm
H}[n]$ and $E_{\rm xc}[n_{\uparrow},n_{\downarrow}]$ are the external potential
and the Hartree and exchange-correlation functionals
respectively\cite{ReviewDFT}.  Atomic units are used throughout the paper.
Lagrange multipliers $\lambda_{ij}^{\sigma}$ are included to impose the
wavefunction orthonormalization condition.  The ground state energy is obtained
by minimizing the total energy functional, namely by imposing $\frac{\delta
F[\{\psi_{i\sigma}\}]}{\delta \langle \psi_{i\sigma}|} = 0$, for each $i$ and
$\sigma$.  Upon derivation of the total energy functional respect to
$\langle\psi_{i\sigma}|$, we obtain:
\begin{eqnarray}
\frac{\delta F}{\delta \langle\psi_{i\sigma}|}=
\left[-\frac{\Delta}{2}+V_{\rm KS}^{\sigma}\right]|\psi_{i\sigma}\rangle+
\sum_{j=1}^{N_{\sigma}} \lambda_{ij}^{\sigma}|\psi_{j\sigma}\rangle
\end{eqnarray} 
where $V_{\rm KS}^{\sigma}({\bf r})=\frac{\delta E_{\rm
KS}[n_{\uparrow},n_{\downarrow}]}{\delta n_{\sigma}({\bf r})}$.

The Hartree functional $E_{\rm H}$ in $F[\{\psi_{i\sigma}\}]$ contains a
self-interaction term for each electron. If the exact functional is used, these
terms are  exactly canceled by an identical term (with opposite sign) included
in $E_{\rm xc}$.  When approximate forms of $E_{\rm xc}$ are used, a
self-interaction is introduced in $F[\{\psi_{i\sigma}\}]$, since $E_{\rm xc}$
no longer cancels the self interaction terms in $E_{\rm H}$ exactly.
This self-interaction is unphysical and must be subtracted.
Since there is one spurious self-interaction term per electron, the elimination
of all the terms leads to an orbital-dependent correction in the functional.
The resulting Kohn-Sham equations are not invariant anymore for a unitary
transformation in the subspace of occupied orbitals, and the minimization of the
functional leads to multi-minima problems.  When DFT calculations give a good
description of the system without the spin $1/2$ center (i.e. pure SiO$_2$) the
relevant self-interaction is typically the unpaired electron's
one\cite{Stokbro,Pacchioni}.  As a consequence we propose to subtract the
self-interaction only for the unpaired electron. More specifically, we remove only
the self-interaction term associated with the magnetization density $m(\mathbf{r})$. 
Indeed, using standard DFT functionals, the Kohn-Sham eigenstates are such that
$|\psi_{i\uparrow}\rangle\simeq |\psi_{i\downarrow}\rangle$ and $m({\bf
r})\simeq |\langle{\bf r}|\psi_{N+1\uparrow}\rangle|^2$ (for which eigenstates
we have $\lambda_{ij}^{\sigma}=\delta_{ij}\epsilon_{i\sigma}$, where
$\epsilon_{i\sigma}$ are the Kohn and Sham eigenvalues).  As it will be shown
below, this strategy eliminates both the orbital dependence of the equations
and the multi-minima problem.

A straightforward way to subtract, at least partially, 
the self-interaction was developed by Perdew and Zunger (SPZ) \cite{Perdew}.
Applying this approach to the unpaired-electron leads to the definition of a
new self-interaction corrected functional (denoted SPZ),
$F_{SIC}[\{\psi_{i\sigma}\}]= F[\{\psi_{\sigma}\}]+\Delta
F_{SPZ}[n_\uparrow,n_\downarrow]$, where
\begin{equation}
\Delta F_{SPZ}[n_\uparrow,n_\downarrow]= -E_{\rm H}[m]-E_{\rm xc}[m,0]
\label{eq:PZ}
\end{equation}
This amounts to subtracting the self-interaction term associated with the
magnetization density from both the Hartree and from the exchange correlation
functionals.

Besides the SPZ scheme, in this work we propose a new SIC functional (denoted
US), defined as $F_{SIC}[\{\psi_{i\sigma}\}]= F[\{\psi_{i\sigma}\}]+ \Delta
F_{US}$, where
\begin{equation}
\Delta F_{US}[n_\uparrow,n_\downarrow]=
-E_{\rm H}[m]-E_{\rm xc}[n_{\uparrow},n_{\downarrow}] 
+E_{\rm xc}[n_{\uparrow}-m,n_{\downarrow}]
\label{eq:SIC}
\end{equation}
namely, (i) we subtract the unpaired electron self-interaction from the Hartree
functional and (ii) we replace the exchange correlation functional for the 2N+1
electrons system with the one for the 2N electrons system without the unpaired electron.

Unfortunately, the US or PZ self-interaction corrected functionals are not
sufficient by themselves to obtain physically relevant densities.  Due to the quadratic
dependence (with a negative sign) of $E_{\rm H}$ respect to $m({\bf r})$, the two
functionals tend to maximize everywhere $|m({\bf r})|$ by separating the spin
up and spin down densities.  As a consequence  $|\psi_{i\uparrow}\rangle$
becomes very different from $|\psi_{i\downarrow}\rangle$ for $i\le N$ and
$m({\bf r})$ is not approximately equal to $|\langle{\bf
r}|\psi_{N+1\uparrow}\rangle|^2$.  This unphysical solution can be eliminated
by introducing a second constraint on the 2N-electrons system, i.e. the system
without unpaired electron.  We impose that up and down electrons with the same
orbital quantum numbers have identical spatial wavefunction (spin-restricted
solution):
\begin{eqnarray}
&|\psi_{i,\uparrow}\rangle &=|\psi_{i,\downarrow}\rangle=|\psi_{i}\rangle
\;\;\;\;\;\;{\rm for}\;\;\; i=1,...,N \label{eq:spinres1}\\
&|\psi_{N+1\uparrow}\rangle &=|\psi_{N+1}\rangle  \label{eq:spinres2}
\end{eqnarray} 
In this way the total energy becomes a function of $\{\psi_{i}\}$: 

\begin{eqnarray}
F_{\rm SIC}[\{\psi_{i}\}]=T[\{\psi_{i\sigma}\}]+E_{\rm KS}[n_{\uparrow},n_{\downarrow}]+ \nonumber \\
+\Delta F_{\rm SIC}[n_\uparrow,n_\downarrow] +
 \sum_{i,j=1}^{N+1} \eta_{ij}
(\langle\psi_i|\psi_j\rangle-\delta_{ij}) \label{eq:functsres}
\end{eqnarray}
where the $\eta_{ij}$ Lagrange multipliers are used to enforce the
orthonormalization conditions.
$\Delta F_{\rm SIC}[n_\uparrow,n_\downarrow]$ is the SIC correction in the SPZ
(eq. \ref{eq:PZ}) or in the US (eq. \ref{eq:SIC}) scheme.  Minimization of the
total energy functional can be achieved by imposing $\frac{\delta F}{\delta
\langle\psi_{i}|}=0$ for $i=1,...,N+1$.  Functional derivation respect to
$\langle\psi_{i}|$ leads to:
\begin{eqnarray}
\frac{\delta F_{\rm SIC}}{\delta \langle\psi_{N+1}|}&=&\left[-\frac{\Delta}{2}+V_{\rm SIC}^{\uparrow}\right]|\psi_{N+1}\rangle+\sum_{j=1}^{N+1}\eta_{N+1 j}|\psi_{j}\rangle \label{eq:unpair}\\ 
\frac{\delta F_{\rm SIC}}{\delta \langle\psi_{i\le N}|}&=&2\left[-\frac{\Delta}{2}+\frac{V_{\rm SIC}^{\uparrow}+V_{\rm SIC}^{\downarrow}}{2}\right]|\psi_i\rangle+\sum_{j=1}^{N+1}\eta_{ij}|\psi_{j}\rangle\nonumber  \label{eq:pair}\\
\end{eqnarray}
The potential $V_{\rm SIC}^{\sigma}$  is is defined as $V_{\rm
SIC}^{\sigma}({\bf r})=V_{\rm KS}^{\sigma}({\bf r})+\frac{\delta \Delta F_{\rm
SIC}}{\delta n_{\sigma}({\bf r})}$, where $\Delta F_{\rm SIC}$ is $\Delta
F_{\rm SPZ}$ or $\Delta F_{\rm US}$ depending on which subtraction scheme is
adopted (eq. \ref{eq:PZ} or eq. \ref{eq:SIC}).

Equations \ref{eq:unpair} and \ref{eq:pair} form a set of self-consistent
equations coupled via the terms involving Lagrange multipliers. The condition
of $\frac{\delta F_{\rm SIC}}{\delta \langle\psi_{i}|}=0$ is equivalent to the
minimization of the corresponding functional.  For fixed ionic positions we
minimize the SPZ and US functionals with the constraints in eq.
\ref{eq:spinres1},\ref{eq:spinres2} using the the gradients given in eq. 
\ref{eq:unpair} and \ref{eq:pair} and the Car-Parrinello\cite{Parrix,FPMD}
method with a damped molecular dynamics approach\cite{Tassone}. 
We used combined electronic and ionic dynamics for the geometry optimizations. 

We simulate the structure of an Al defect in Silica using a neutral cell of 72
atoms (one Al atom, 23 Si and 48 O).  We perform electronic structure
calculations \cite{FPMD,Paratec} using DFT in the spin polarized generalized
gradient approximation and the Perdew-Burke-Ernzerhof (PBE) functional
\cite{PBE} corrected for self-interaction as in eq. \ref{eq:PZ} and in eq.
\ref{eq:SIC}.  We use norm conserving pseudo-potentials \cite{Troullier}.
The wave functions are expanded in plane waves using a $70$ Ry cutoff.  We
sample the Brillouin zone with the $\Gamma$ point and we impose the $\langle
{\bf r}|\psi_i\rangle$ to be real \cite{footnote}.

\begin{table}[h]
\begin{ruledtabular}
\begin{tabular}{lcc}
Functional & Delocalized geometry & Localized geometry \\ \hline
PBE &  stable  &  unstable (43.9 mRy) \\
SPZ &  stable  &  metastable (1.3 mRy) \\
US  & unstable (238 mRy) & stable \\
UHF & unknown    & stable  \\
\end{tabular}
\end{ruledtabular}
\caption{\label{tab:struct}Stability of of localized and delocalized geometries
with different functionals. In parenthesis we indicate the energy difference
between the metastable/unstable structure and the stable one.  With PBE  we use
the US minimal energy structure as the localized geometry.  With US  we use the
PBE minimal energy structure as the delocalized geometry.}
\end{table}

\begin{table}[h]
\begin{ruledtabular}
\begin{tabular}{lccccc}
  &\multicolumn{2}{c}{Delocalized geometries}&\multicolumn{3}{c}{Localized geometries} \\ \hline
Bond      & PBE\cite{PBE} &  SPZ   & SPZ   & US    & UHF\cite{Stokbro} \\ \hline
Al-O(1)  & 1.744         &  1.742   & 1.938 & 1.957 & 1.924  \\
Al-O(2)  & 1.744         &  1.742   & 1.713 & 1.710 & 1.688  \\
Al-O(3)  & 1.753         &  1.748   & 1.706 & 1.705 & 1.703  \\
Al-O(4)  & 1.753         &  1.748   & 1.719 & 1.715 & 1.689  \\ \hline
Si-O(1)   & 1.613        &  1.596   & 1.756 & 1.793 & - \\
Si-O(2)   & 1.613        &  1.596   & 1.604 & 1.602 & - \\
Si-O(3)   & 1.618        &  1.595   & 1.595 & 1.595 & - \\
Si-O(4)   & 1.618        &  1.595   & 1.601 & 1.601 & - \\
\end{tabular}
\end{ruledtabular}
\caption{\label{tab:bondsAl}Bond-lengths ($\rm{\AA}$) around the substitutional Al
impurity using unrestricted Hartree-Fock and density functional theory with
functionals PBE\cite{PBE},SPZ and US.}
\end{table}

The results of geometrical optimization using different functionals are
illustrated in tables \ref{tab:struct} and \ref{tab:bondsAl}.  We label
{\it delocalized geometry} a structure in which the hole wavefunction is
delocalized on the four O atoms surrounding the Al one.  On the contrary a
{\it localized geometry} is a structure in which the hole wavefunction is
localized on one of the four surrounding Oxygens.  For a given functional we
indicate the structure corresponding to an absolute minimum with {\it stable} and to
a local minimum (but not the absolute one) with {\it metastable}.  The structure not
corresponding to any minimum is called {\it unstable}.  The PBE stable
structure\cite{Stokbro} is a delocalized geometry, in qualitative disagreement
with UHF.  The SPZ stable structure is delocalized.  However a metastable
localized solution occurs at slightly higher energy (1.3 mRyd).  The US stable
structure is localized, in agreement with the UHF results. We did not find any
metastable solution using the US functional and in particular the delocalized
structures found with PBE or SPZ are unstable.
Finally, to judge the effects of the spin-restricted prescription, we compute
the total energy difference between the localized and the delocalized
structures using PBE with the constraint of eqs. \ref{eq:spinres1} and
\ref{eq:spinres2}.  We obtain 45.6 mRyd which is very close to the unrestricted
result of 43.9 mRyd given in table \ref{tab:struct}, i.e. the spin-restricted
condition weakly affects the total energy differences.

Delocalized geometries are tetrahedral structures with Al-O and Si-O
bond-lengths very close to the Al-O distance ($\approx 1.73 {\rm \AA}$) in
AlPO$_4$ and to the Si-O bond-lengths in quartz ($\approx 1.61 {\rm \AA}$),
respectively.  The localization of the hole on one particular oxygen (labeled
O(1) in tab.  \ref{tab:bondsAl}) leads to a distorted tetrahedral structure
with two elongated Al-O(1) and Si-O(1) bonds, while all the others Al-O bonds
are only slightly smaller than in the delocalized case.  The   Al-O and Si-O
bond-lengths of stable US and metastable SPZ structures are in good agreement
with UHF. 

\begin{table}[h]
\begin{ruledtabular}
\begin{tabular}{ccccccc}
Geom:& PBE   & US       & SPZ     & US  & UHF\cite{Stokbro} & \\
Fun:& PBE    & PBE      & SPZ     & US  & UHF\cite{Stokbro} & Exp.\\\hline
    &-17.0  & -72.9  & -96.9 & -89.5 & -89.3 &-85.0 \\
O(1)&  8.46 &  36.3  &  48.3 &  44.6 &  44.6 & 41.2 \\
    &  8.55 &  36.7  &  48.6 &  44.8 &  44.7 & 43.8 \\
\end{tabular}
\end{ruledtabular}
\caption{\label{tab:hyperfine} Anisotropic hyperfine parameters (Gauss) of the
$^{17}$O (1) atom. The first row ``Geom:'' indicates the functional used to
determine the geometry, the second row ``Fun:'' the functional used in the
hyperfine coupling calculation.  In this table SPZ geometry refers to the
metastable localized structure obtained with the SPZ functional (see text). }
\end{table}

To validate our SIC methods against experiment  we compute hyperfine parameters
using the formalism developed in ref. \cite{VandeWalle}. The hyperfine
interaction includes an isotropic part (Fermi contact) and an anisotropic
(dipolar) part.  The isotropic part measures the spin density at the 
nucleus and it is non-zero only for wavefunctions containing s-wave components
\cite{VandeWalle}.  The hole state, both in the 
localized and delocalized solutions, corresponds to the O 2p lone-pair state.
Thus the Fermi contact is mainly due to the spin polarization of the doubly
occupied O 2s state.  Since the spin-restricted conditions (eqs.
\ref{eq:spinres1} and \ref{eq:spinres2}) suppresses such spin polarization, the
Fermi contact term cannot be computed with our spin-restricted SIC functionals
\cite{footnote2}.  On the other hand, we can access the anisotropic terms since
they capture the $p-$like component of the electron wavefunction which is
weakly affected by the spin-restricted condition \cite{footnote2}.  In table
\ref{tab:hyperfine} we report the principal values of the dipolar part for
$^{17}$O  atoms.  Our implementation of the SIC functional substantially
improves the PBE results, giving a dipolar part very close to the UHF
results and the experimental data. 

In this work we have presented a very effective and  easy-to-implement
self-interaction corrected approach. It can be applied to hole or electron
spin-1/2-centers in solids.  We have validate our
method by calculating energetics, geometries and hyperfine couplings of neutral
Al substitutional defects in quartz. Our SIC approach corrects the known
deficiencies\cite{Stokbro,Pacchioni} of standard DFT functionals and gives
results in good agreement with experiments and with the self-interaction free
UHF calculations.  Thus the functional proposed in this work solves the
problem\cite{Stokbro} of describing the behavior of hole self-trapping in
silica in the framework of density functional theory. Finally we note that our
approach can also be applied to molecular systems such as spin 1/2 radicals
which represents a fundamental subject of research in chemistry and
biochemistry.

We acknowledge illuminating discussions with C. Cavazzoni.  The calculations
were performed at the IDRIS supercomputing center (project 031202).  M. C. was
supported by Marie Curie Fellowships under Contract No. IHP-HPMF-CT-2001-01185.


\begin{thebibliography}{99}
\bibitem{footnote} 
 In absence of an external magnetic field the manybody 
wavefunction is real. 
However using the SPZ functional (but not the US functional)
with a complex wavefunction we found a localized solution having
 32.1mRy lower energy than that obtained with a real wavefunction.
Such complex solution is non-physical.
For instance, its hyperfine dipolar principal values are non-axially symmetric
[(-49.0, 11.6, 37.4) Gauss] in stark contrast with the experimental findings and
with the fact that a p-like real wavefunction should lead to an 
axially symmetric dipolar part.
\bibitem{footnote2} The O(1) Fermi contact
interaction in the US geometry is -39.5 Gauss with PBE while 
we get 0.004 Gauss with the same functional but imposing the
spin-restricted condition. The corresponding anisotropic 
principal values are (-72.9, 36.3, 36.7) Gauss and 
(-66.8, 33.3, 33.5) Gauss, respectively.
\end{thebibliography}
\end{document}